\documentclass[12pt,epsf]{article}
\usepackage{amsmath,amssymb}
\usepackage{here}
\usepackage{graphicx}
\usepackage{cancel}
\def\slashchar#1{\setbox0=\hbox{$#1$} 
\dimen0=\wd0 
\setbox1=\hbox{/} \dimen1=\wd1 
\ifdim\dimen0>\dimen1 
\rlap{\hbox to \dimen0{\hfil/\hfil}} 
#1 
\else 
\rlap{\hbox to \dimen1{\hfil$#1$\hfil}} 
/ 
\fi}

\begin{document}

\def\a{\alpha}
\def\b{\beta}
\def\c{\varepsilon}
\def\d{\delta}
\def\e{\epsilon}
\def\f{\phi}
\def\g{\gamma}
\def\h{\theta}
\def\k{\kappa}
\def\l{\lambda}
\def\m{\mu}
\def\n{\nu}
\def\p{\psi}
\def\q{\partial}
\def\r{\rho}
\def\s{\sigma}
\def\t{\tau}
\def\u{\upsilon}
\def\v{\varphi}
\def\w{\omega}
\def\x{\xi}
\def\y{\eta}
\def\z{\zeta}
\def\D{\Delta}
\def\G{\Gamma}
\def\H{\Theta}
\def\L{\Lambda}
\def\F{\Phi}
\def\P{\Psi}
\def\S{\Sigma}

\def\o{\over}
\def\beq{\begin{eqnarray}}
\def\eeq{\end{eqnarray}}
\newcommand{\gsim}{ \mathop{}_{\textstyle \sim}^{\textstyle >} }
\newcommand{\lsim}{ \mathop{}_{\textstyle \sim}^{\textstyle <} }
\newcommand{\vev}[1]{ \left\langle {#1} \right\rangle }
\newcommand{\bra}[1]{ \langle {#1} | }
\newcommand{\ket}[1]{ | {#1} \rangle }
\newcommand{\EV}{ {\rm eV} }
\newcommand{\KEV}{ {\rm keV} }
\newcommand{\MEV}{ {\rm MeV} }
\newcommand{\GEV}{ {\rm GeV} }
\newcommand{\TEV}{ {\rm TeV} }
\def\diag{\mathop{\rm diag}\nolimits}
\def\Spin{\mathop{\rm Spin}}
\def\SO{\mathop{\rm SO}}
\def\O{\mathop{\rm O}}
\def\SU{\mathop{\rm SU}}
\def\U{\mathop{\rm U}}
\def\Sp{\mathop{\rm Sp}}
\def\SL{\mathop{\rm SL}}
\def\tr{\mathop{\rm tr}}

\def\IJMP{Int.~J.~Mod.~Phys. }
\def\MPL{Mod.~Phys.~Lett. }
\def\NP{Nucl.~Phys. }
\def\PL{Phys.~Lett. }
\def\PR{Phys.~Rev. }
\def\PRL{Phys.~Rev.~Lett. }
\def\PTP{Prog.~Theor.~Phys. }
\def\ZP{Z.~Phys. }


\baselineskip 0.7cm

\begin{titlepage}

\begin{flushright}
IPMU14-0052 \\
UT-14-8
\end{flushright}

\vskip 1.35cm
\begin{center}
{\large \bf Testing the Minimal Direct Gauge Mediation at the LHC
}
\vskip 1.2cm
  {\bf   Koichi Hamaguchi$^{(1)(2)}$, Masahiro Ibe$^{(2)(3)}$, Tsutomu T. Yanagida$^{(2)}$\\ and Norimi Yokozaki$^{(2)}$}
\vskip 0.4cm

{\it  
$^{(1)}$Department of Physics, University of Tokyo, Bunkyo-ku, Tokyo 113-0033, Japan \\
$^{(2)}$Kavli IPMU (WPI), TODIAS, University of Tokyo, Kashiwa 277-8583, Japan\\
$^{(3)}$ICRR, University of Tokyo, Kashiwa 277-8582, Japan\\
}

\vskip 1.5cm

\abstract{
We reexamine the models with gauge mediation in view of the minimality.
As a result, we arrive at a very simple model of direct gauge mediation 
which does not suffer from the flavor problems nor the CP problems.
We show that the parameter space  which is consistent with the Higgs boson mass at around $126$\,GeV 
can be tested 
through the stable stau searches at the 14TeV run of the LHC. The gravitino is a viable candidate for a dark matter. We also give a short discussion on a possible connection of our model
to the recently discovered X-ray line signal at 3.5\,keV in the XMM Newton X-ray observatory data.
 
}

\end{center}
\end{titlepage}

\setcounter{page}{2}

\section{Introduction}

The minimal supersymmetric Standard Model (MSSM) has been widely believed to be one of the
most attractive models of physics beyond the Standard Model (SM), since it provides a solution
to the hierarchy problem between the scale of the SM and the very high energy scales 
such as the Planck scale or the scale of the Grand Unified Theory (GUT).
The precise unification of the gauge coupling constants at the GUT scale has also supported
the MSSM  very strongly.

The observed Higgs boson mass at around 126 GeV\,\cite{Higgs} is, however, near the upper limit of
the predictions in most conventional models of the MSSM with SUSY particle masses 
in hundreds GeV to a few TeV range.
This rather large Higgs boson mass and the so far null results of the SUSY particle
searches at the LHC have stirred up fears of non discovery of SUSY particles even at 
the upgraded 14\,TeV run of the LHC.


In this paper, we reexamine the MSSM with fresh eyes, by placing 
greater emphasis on minimality  as a guiding principle for model building 
rather than the conventional naturalness.
In the course of the application of  minimality, we take the models with gauge 
mediation\,\cite{Dine:1981za, Dine:1981gu, Dimopoulos:1982gm,Affleck:1984xz,
Dine:1993yw,Dine:1994vc,Dine:1995ag}
 as our starting point,
since it solves the SUSY flavor changing neutral current (FCNC) problem in the minimal set up.
In particular, we confine ourselves to the models with direct gauge mediation in line with minimality.
We further shave the models by assuming that the  $\mu$-term is given just as it is.
As a result, we  end up with models with very few parameters which are  free from not only the SUSY FCNC problem 
but also the SUSY CP problem.
We call this minimal model as the minimal direct gauge mediation. 

As we will see, the squark and the gluino masses are required to be in the multi-TeV range to explain 
the Higgs boson mass at around 126\,GeV, which are beyond the reach of the 14\,TeV run.
Interestingly, however, we find the stau is predicted to be rather light due to a large tau 
Yukawa coupling constant in the this model.
As a result, we find that there is a large parameter region where the stau is the next-to the lightest particle
(NLSP) with a mass below 1.0--1.2\,TeV.
We also find that the mass of the gravitino is typically above the keV range, and hence,
the stau can be long lived and decay outside the LHC detectors, which provides 
searchable signals at the 14\,TeV run of the LHC. We also discuss a possible connection of our model to the recently discovered X-ray line signal at 3.5\,keV in the XMM Newton X-ray observatory data; if the model is embedded into string theories, a moduli has a similar mass to the gravitino mass which can be a dark matter.

The organization of the paper is as follows.
In section\,\ref{sec:minimalmodel}, 
we reconsider the models with gauge mediation in view of minimality.
In section\,\ref{sec:predictions}, we show the predictions of the minimal
direct gauge mediation model.
There, we find that the stau can be the NLSP and decay outside the detectors of the LHC
depending on the gravitino mass.
The final section is devoted to discussions and conclusions.

\section{Putting Minimality  on Gauge Mediation}\label{sec:minimalmodel}
The minimal ingredients of  the models with gauge mediation are a sector of spontaneous SUSY 
breaking and a sector of the messenger fields.
Here, 
we collectively represent the SUSY breaking sector in terms of a single SUSY breaking field $Z$ whose
Lagrangean is given by, 
\begin{eqnarray}
{\cal L}_{\cancel{\rm SUSY}} 
 =  \int d^4 \theta \left[ Z^\dag Z - \frac{(Z^\dag Z)^2}{4\Lambda^2}\right] + \int d^2 \theta  \left[ -\mu_Z^2 Z \right] + h.c.
\end{eqnarray}
Here, $\mu_Z$ and $\Lambda$ are dimensionful parameters which are determined by the dynamics in the SUSY
breaking sector.%
\footnote{Although the detail structure of the SUSY breaking sector is not relevant 
for the following discussions, we may consider concrete models of SUSY breaking
such as O'Raifeartaigh model\,\cite{O'Raifeartaigh:1975pr} models or 
vector-like dynamical SUSY breaking models\,\cite{Izawa:1996pk} 
behind this simplified description.
}
In this simplified description of the SUSY breaking sector, SUSY is spontaneously broken by the vacuum expectation value (vev) 
of the $F$-term of $Z$,%
\begin{eqnarray}
\label{eq:SUSYv}
\vev{Z} = 0 \ ,\quad F = \vev{F_Z} = \mu_Z^2\ ,
\end{eqnarray}
while the pseudo-flat direction obtains a mass from the quartic coupling in the K\"ahler potential; 
\begin{eqnarray}
m_Z^2 \simeq \frac{|F|^2}{\Lambda^2}\ .
\end{eqnarray}

In choosing the messenger fields, we assume that the messenger 
fields do not take part in dynamics of the SUSY breaking sector and  are simply pairs
of some representations and  anti-representations of the minimal GUT group $SU(5)$, $(\Psi,\bar{\Psi})$
to keep the minimality of the model.
We further assume that the messenger fields coupled to the SUSY breaking sector directly through 
the Yukawa interactions,
\begin{eqnarray}
W = k Z \bar\Psi \Psi + M_{\rm mess} \bar\Psi \Psi\ , 
\end{eqnarray}
where $k$ is a dimensionless coupling constant.
Here, we have also introduced an explicit mass term for the messenger fields, $M_{\rm mess}$,
so that we can avoid extending the models to generate non-vanishing vev of the $A$-component of $Z$\,\cite{Dine:1993yw,Dine:1994vc,Dine:1995ag,Shih:2007av,Komargodski:2009jf,Ibe:2010jb,Evans:2011pz,Curtin:2012yu}.\footnote{
It should be noted that the above direct coupling between the messenger fields and the SUSY breaking field makes
the SUSY breaking vacuum in Eq.\,(\ref{eq:SUSYv}) meta-stable, which leads to a constraint~\cite{Hisano:2008sy}. 
However, this constraint is avoided in the relevant region discussed in the next section.}
With this minimal set up, the MSSM gauginos and the scalars obtain the soft masses
via gauge mediation at the messenger mass scale,
\begin{eqnarray}
M_{\rm gaugino} \simeq \frac{g^2}{16\pi^2} N_{\rm mess} \frac{k \mu_Z^2}{M_{\rm mess}}\ , 
\ m_{\rm scalar}^2 \simeq \frac{2C_2 g^4}{(16\pi^2)^2}  N_{\rm mess} \left|\frac{k \mu_Z^2}{M_{\rm mess}}\right|^2\ .
\end{eqnarray}
Here, $g$ collectively represents the gauge coupling constants of the MSSM, 
$C_2$ is a quadratic Casimir ($C_2 = (N^2-1)/(2N)$ for $SU(N)$), and 
$N_{\rm mess}$ denotes the effective number of the pairs of the messenger fields
in terms of the fundamental representation of $SU(5)$.

Finally, let us discuss the origin of the $\mu$-term.
As is well known, it is the long-sought problem to provide 
the $\mu$ and $B\mu$-terms which are in the similar size of the other soft parameters.
Here,  giving priority to the minimality again, we assume that the $\mu$-term is given just as it is, i.e.
\begin{eqnarray}
W_{\mu} = \mu H_u H_d\ .
\end{eqnarray}
The notable feature of this type of the $\mu$-term is that the gauge mediated $B$-term is 
vanishing at around the messenger scale at the one-loop level,%
\footnote{
For a threshold corrections to $B$ from the two-loop contributions, see Refs.\,\cite{Rattazzi:1996fb,Kahn:2013pfa}.
}
\begin{eqnarray}
B \simeq 0\ .
\end{eqnarray}
As we will see in the next section, this boundary condition plays a very important role in
narrowing down the predictions of the model.%
\footnote{See Refs.\,\cite{Rattazzi:1996fb,Gabrielli:1997jp,Hisano:2007ah} for earlier studies 
on models with $B = 0$ at the messenger scale.}

In summary of the minimal set up of the models with gauge mediation, we arrive at a simple model; 
\begin{eqnarray}
\label{eq:minimal}
\mathcal{L} = \int d^4 \theta \left[ Z^\dag Z - \frac{(Z^\dag Z)^2}{4\Lambda^2}\right] + \int d^2 \theta  \left[ -\mu_Z^2 Z + (kZ+M_{\rm mess}) \bar{\Psi} \Psi +\mu H_u H_d \right]\ .
\end{eqnarray}
We call this model the minimal direct gauge mediation (MDGM) model.%
\footnote{
For a messengers of a pair of ${\bf 5}$ + ${\bf \bar{5}}$, there can be a mixing term between the Higgs doublets 
and the messengers without any additional 
symmetries\,\cite{Chacko:2001km,Chacko:2002et,Evans:2011bea,Evans:2012hg,Craig:2012xp,Abdullah:2012tq}. 
When the messenger sector consists of a pair of ${\bf 10}$ + ${\bf \bar{10}}$ representations of $SU(5)$, on the contrary,
the mixing between the messengers and the Higgs doublets is automatically suppressed.
}
One of the most important advantage of this minimal model is that all the complex phases of the parameters 
 in Eq.\,(\ref{eq:minimal}) can be rotated away by the phase rotations of $Z$, $\Psi,\bar{\Psi}$, $H_{u,d}$
and the superspace coordinate $\theta$.%
\footnote{
When there are multiple messengers, we cannot rotate away all the phases in the parameters.
Even in that case, the phases of the generated gaugino masses in the MSSM are common, 
and hence, we can rotate away all the phases from the MSSM soft terms.
}
Therefore, the MDGM model is not only free from the SUSY FCNC problems 
but also free from the SUSY CP problems.%
\footnote{
CP violation from the supergravity mediated effect
 will be discussed in the next section.}

As will be studied in the next section, the MSSM spectrum is determined by 
only three parameters,
\begin{eqnarray}
M_{\rm eff} = \frac{k\mu_Z^2}{M_{\rm mess}}
 \ , \quad M_{\rm mess}\ , \quad N_{\rm mess} \ ,
\end{eqnarray}
while $\mu$ is fixed by the  electroweak symmetry breaking (EWSB) conditions.
It should be also noted that the EWSB conditions result in a large $\tan\beta$ 
due to the vanishing $B$-term at the messenger scale, 
which leads the stau NLSP in a large parameter space.

One more important parameter for the LHC phenomenology is the 
gravitino mass, 
\begin{eqnarray}
m_{3/2} = \frac{F}{\sqrt{3}M_{P}} = \frac{\mu_Z^2}{\sqrt{3}M_P}\ ,
\end{eqnarray}
which determines the lifetime of the NLSP (see appendix\,\ref{sec:lifetime}). 
Here, $M_P$ denotes the reduced Planck scale, $M_P \simeq 2.4 \times 10^{18}$\,GeV.
As we will show in the next section, the Higgs boson mass at around $126$\,GeV
can be explained in the parameter space where the NLSP mass is at around $1$\,TeV.
For $m_{3/2} > O(100)$\,keV, the NLSP decays outside the detectors,  
and hence, we can detect the SUSY events by searching for charged tracks when
the stau is the NLSP.

\section{Minima Direct Gauge Mediation at the LHC}\label{sec:predictions}
In the MDGM model, the right-handed stau becomes lighter than the neutralino in a large parameter space, 
due to the {\it predicted} large $\tan\beta$. 
To see how large $\tan\beta$ is predicted, let us estimate the $B$-term at the weak scale which 
is mainly generated from the gaugino masses through renormalization group (RG) evolution; 
\beq
\frac{d B}{d \ln Q} \simeq \frac{1}{8\pi^2}(3 g_2^2 M_2 + \frac{3}{5} g_1^2 M_1 + 3 Y_t^2 A_t + 3Y_b^2 A_b  + Y_{\tau}^2 A_{\tau})\ ,
\eeq
with the boundary condition of $B(Q=M_{\rm mess})=0$ ($Q$ is a renormalization scale). 
Here, we denote $M_2$ and $M_1$ as the Wino mass and Bino mass, respectively. 
The Yukawa coupling constants and the scalar trilinear couplings of the top (stop), the bottom (sbottom) and the tau (stau) 
are denoted by $Y_t$, $Y_b$, $Y_\tau$,  $A_t$, $A_b$ and $A_{\tau}$, respectively.

Since the $B$-term generated by the RG evolution is small even at the weak scale, $\tan\beta$ is predicted to be large 
after imposing the EWSB conditions;
\beq
\frac{m_Z^2}{2} &\simeq& \frac{(m_{H_d}^2 + \frac{1}{2v_d}\frac{\partial\Delta V}{\partial v_d}) - (m_{H_u}^2 +\frac{1}{2v_u}\frac{\partial\Delta V}{\partial v_u})\tan^2\beta }{\tan^2 \beta -1} - \mu^2\ , \label{eq:ewsb_A}\\
 B\mu (\tan\beta + \cot \beta) &\simeq& \left(m_{H_u}^2 + \frac{1}{2v_d}\frac{\partial\Delta V}{\partial v_u}+ m_{H_d}^2 +\frac{1}{2v_u}\frac{\partial\Delta V}{\partial v_u}+ 2 \mu^2 \right)  \label{eq:ewsb_B}\ ,
\eeq
where $v_u =\left<H_u^0\right>$ ($v_d=\left<H_d^0\right>$) is the vacuum expectation value of the up type (down type) Higgs doublets. 
The one-loop corrections to the Higgs potential is denoted by $\Delta V$. 
The Higgs soft masses for the up-type and down-type Higgs are $m_{H_u}^2$ and $m_{H_d}^2$, respectively.
Here, we take the convention that $B \mu$ is real positive. 
Since all the soft masses, $m_{H_u}^2$, $m_{H_d}^2$ and $B$, are fixed for given parameters of gauge mediation, 
$\tan\beta$ as well as $\mu$ are determined by solving Eq.(\ref{eq:ewsb_A}) and (\ref{eq:ewsb_B}).
As a result, the one-order of magnitude smaller $B$-term compared with $(m_{H_d}^2+|\mu|^2)$ leads to a 
large $\tan\beta$ of about $30$--$60$ depending on the messenger scale.

\begin{figure}[t]
\begin{center}
\includegraphics[scale=0.95]{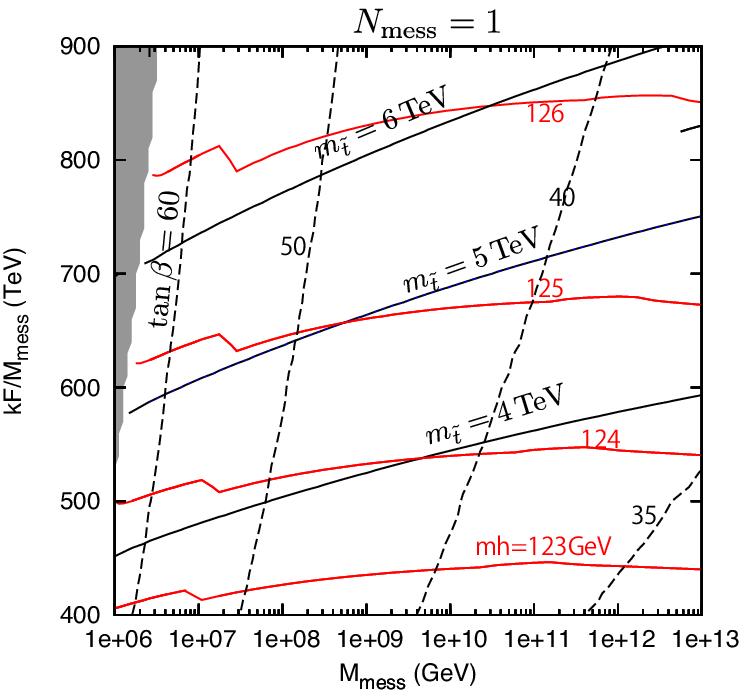}
\includegraphics[scale=0.95]{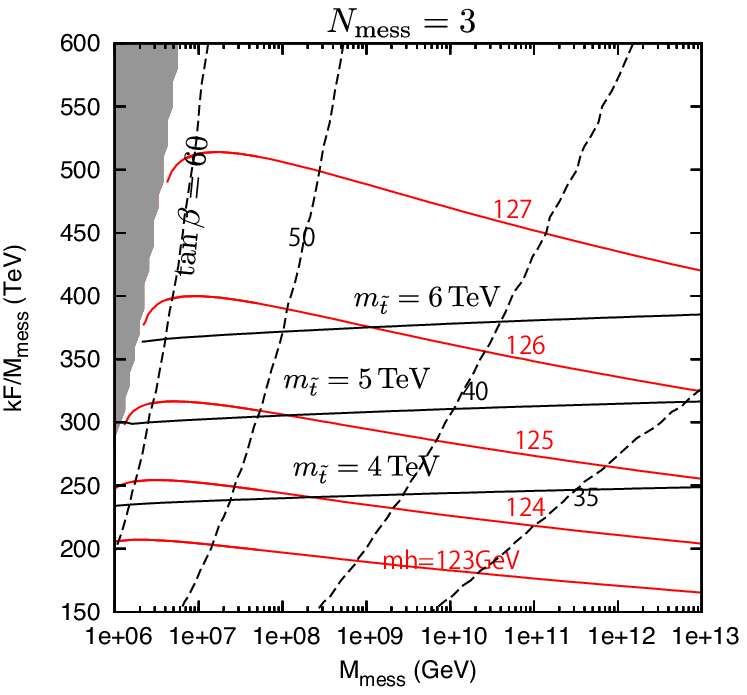}
\caption{\sl \small The lightest Higgs boson mass (red solid line), stop mass ($m_{\tilde{t}}\equiv \sqrt{m_{\tilde{Q}_3} m_{\tilde{U}_3^c}}$) (black solid line) and predicted $\tan\beta$ (black dashed line). Here $m_t=173.3$ GeV and $\alpha_S(m_Z)=0.1184$.
}
\label{fig:stop}
\end{center}
\end{figure}

In Fig.~\ref{fig:stop}, the contours of $\tan\beta$ are shown on $M_{\rm mess}$--$(kF/M_{\rm mess})$ plane with the messenger number $N_{\rm mess}=1$ and $3$. The SUSY mass spectra as well as the renormalization group running are calculated by using the {\tt SOFTSUSY} package~\cite{softsusy}. The larger messenger scale predicts a smaller $\tan\beta$, 
since the generated $B$ becomes larger due to the logarithmic enhancement from the messenger scale to the weak scale. 
In the gray shaded region, the successful electroweak symmetry breaking does not occur because of too large negative $m_{H_d}^2$ with which Eq.\,(\ref{eq:ewsb_B}) can not be satisfied; the predicted $\tan\beta$ is too large and the large bottom/tau Yukawa couplings drive $m_{H_d}^2$ negative  too much.

The Higgs boson mass in MDGM is dominantly raised by the radiative corrections from heavy stops~\cite{OYY}.
Fig.~\ref{fig:stop} also shows the Higgs boson mass and the required value of the stop mass parameter defined 
by $m_{\tilde{t}} \equiv \sqrt{m_{\tilde Q _3} m_{\tilde {\bar U} _3^c}}$ ($m_{\tilde Q _3}$ 
and $m_{\tilde U _3^c}$ are left- and right handed stop mass, respectively). The Higgs boson mass is obtained 
by using {\tt FeynHiggs2.10.0}~\cite{feynhiggs}. 
Since $A_t$ is not large, the stop mass should be large as $m_{\tilde{t}} \sim 5-6$ TeV. Thanks to the large $\tan\beta$ of $\mathcal{O}(10)$, the very heavy stops of $\mathcal{O}(50-100)$ TeV are not required. However, the colored SUSY particles are too heavy to be produced at the LHC even at 14\,TeV run.

\begin{figure}[t]
\begin{center}
\includegraphics[scale=0.7]{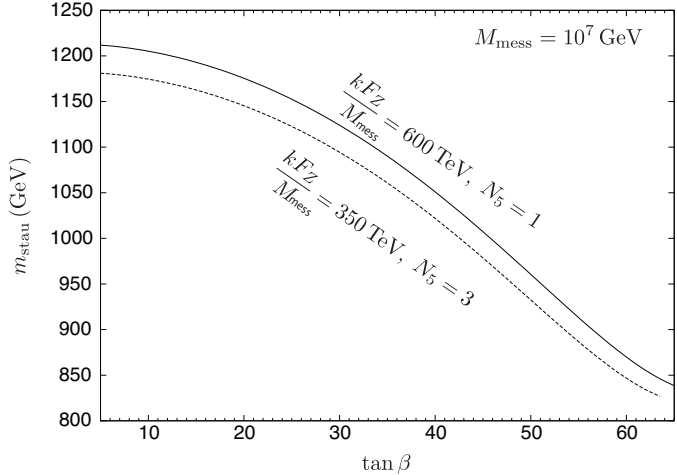}
\caption{\sl The stau mass as a function of $\tan\beta$. Here, we take $\tan\beta$ to be an input parameter, and the Higgs B-parameter is determined by Eq.(13).
}
\label{fig:stau_tanb}
\end{center}
\end{figure}


\begin{figure}
\begin{center}
\includegraphics[scale=0.95]{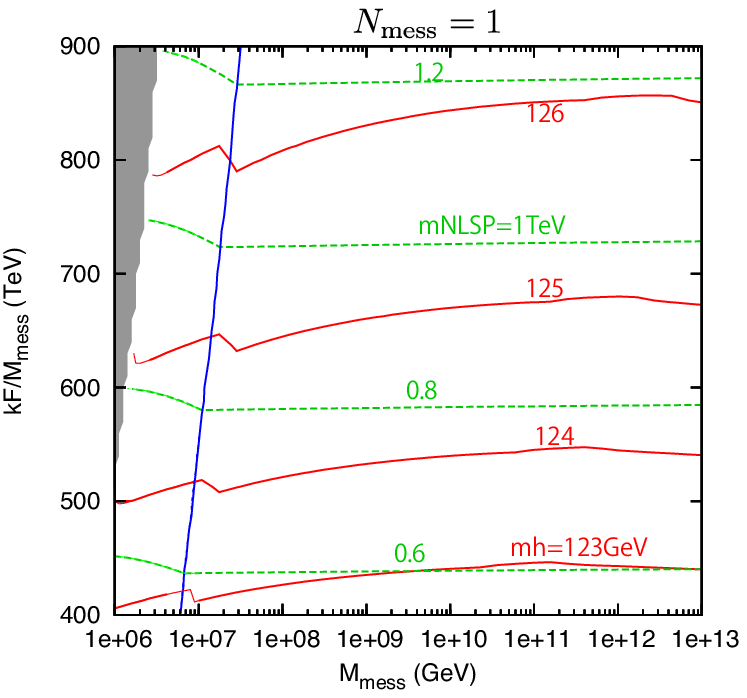}
\includegraphics[scale=0.95]{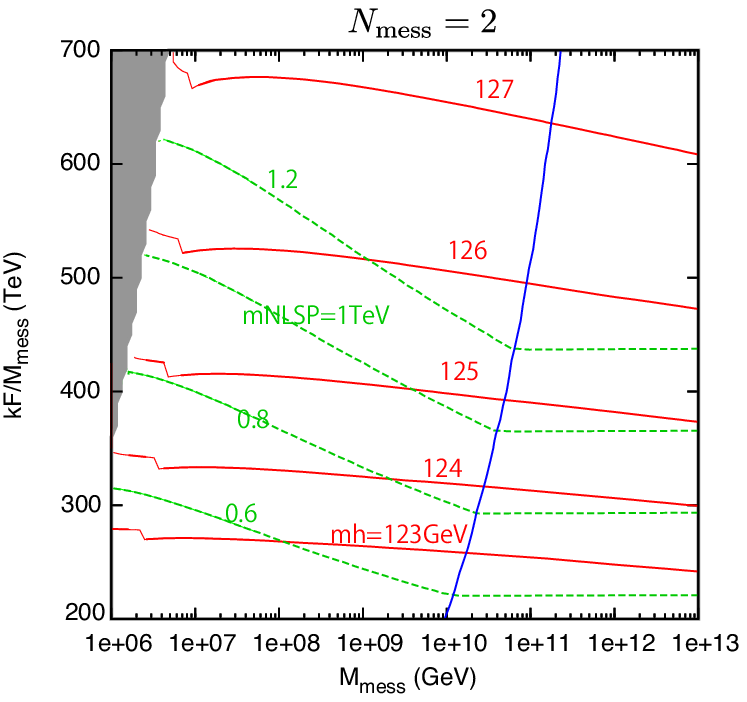}
\includegraphics[scale=0.95]{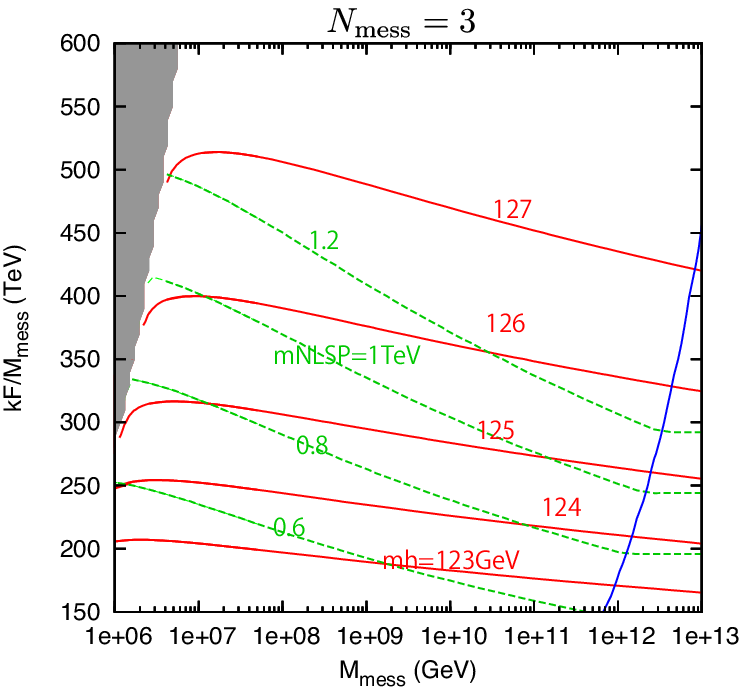}
\includegraphics[scale=0.95]{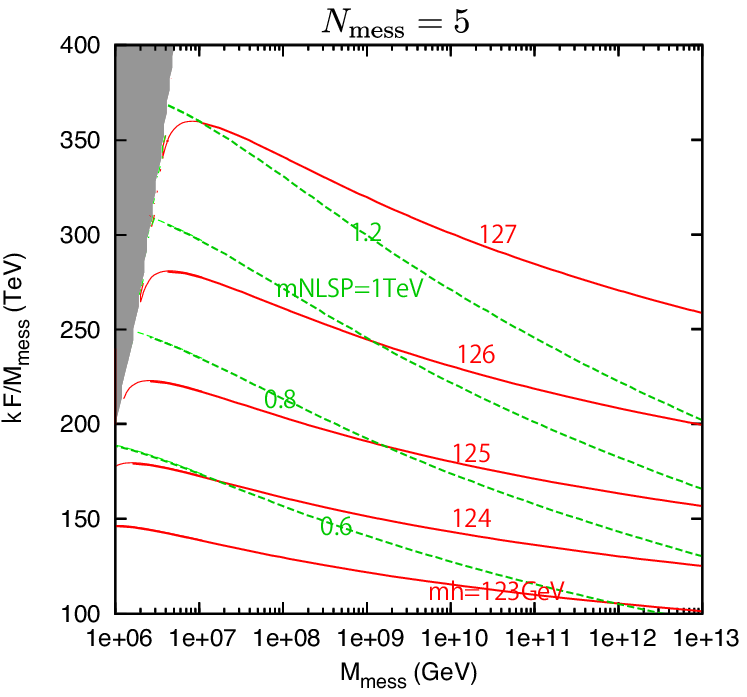}
\caption{\sl \small Contours of the NLSP mass and Higgs boson mass for the different messenger numbers. The Higgs $B$-term is taken as $B(\rm Mess)=0$. In the left (right) region of the blue line, the stau (neutralino) is the NLSP.
}
\label{fig:stau}
\end{center}
\end{figure}

Since the tau Yukawa coupling is enhanced for a large $\tan\b$ by $Y_{\tau} \simeq m_{\tau}/v \tan\beta$, 
the radiative corrections proportional to $Y_{\tau}^2$ is larger than the case with e.g., $\tan\beta \simeq 10$.
In particular, the stau masses receive the RG corrections and threshold corrections which are 
proportional to $Y_{\tau}^2$, 
\beq
\frac{d m_{\tilde{\tau}_R}^2}{dt} \sim \frac{Y_{\tau}^2}{4\pi^2} \left( m_{\tilde{L}_3}^2 + m_{\tilde{\tau}_R}^2 + m_{H_d}^2  + A_{\tau}^2\right) + \dots,
\eeq
and 
\begin{eqnarray}
m_{\tilde{\tau}_R}^2(\tilde{\tau}_R) \hspace{-2.2pc}&&-\, m_{\tilde{\tau}_R}^2(m_{\tilde{t}}^2) \nonumber \\
\hspace{-8pt}&\simeq& \frac{Y_{\tau}^2}{4\pi^2} \left(2 \mu^2 \ln \frac{m_{\tilde{t}}}{\mu} - m_A^2 \ln \frac{m_{\tilde{t}}}{m_A}- \mu^2 \ln \frac{m_{\tilde{t}}}{m_{\tilde{L}_3}} -m_{\tilde{L}_3}^2\ln \frac{m_{\tilde{t}}}{m_{\tilde{L}_3}}-m_{\tilde{\tau}_R}^2\ln \frac{m_{\tilde{t}}}{m_{\tilde{\tau}_R}} \right)\ .
\end{eqnarray}
Both corrections are not negligible for a large $\tan\beta$. In Fig.~\ref{fig:stau}, we show the stau mass as a function of $\tan\beta$. Comparing the stau mass for $\tan\beta=10$ with that for $\tan\beta=60$, it is differed by $\sim$300\,GeV. 
 As a result, we find that  the right-handed stau mass becomes light and the stau can be NLSP even for $N_{\rm mess}=1$.
We also find that the stau becomes the NLSP in a large parameter space for $N_{\rm mess} \geq 2$.

In Fig.~\ref{fig:stau}, contours of the NLSP masses are shown. In the left (right) region of the blue line, the stau (neutralino) is the NLSP. The mass of the NLSP is bounded from above as $m_{\rm NLSP} < 1.0$--$1.2$\,TeV. 
In the case of the stau NLSP with this mass, it can be observed if it is stable inside the detector~\cite{cms_stau}.
Note that the low-energy mass spectra in $N_{\rm mess}=3$ and $5$ are (almost) identical to those induced by a pair of messengers in $\bf 10$ ($\bf \overline{10}$) and $\bf 24$ representation of $SU(5)$, respectively.

\begin{figure}[t]
\begin{center}
\includegraphics[scale=0.95]{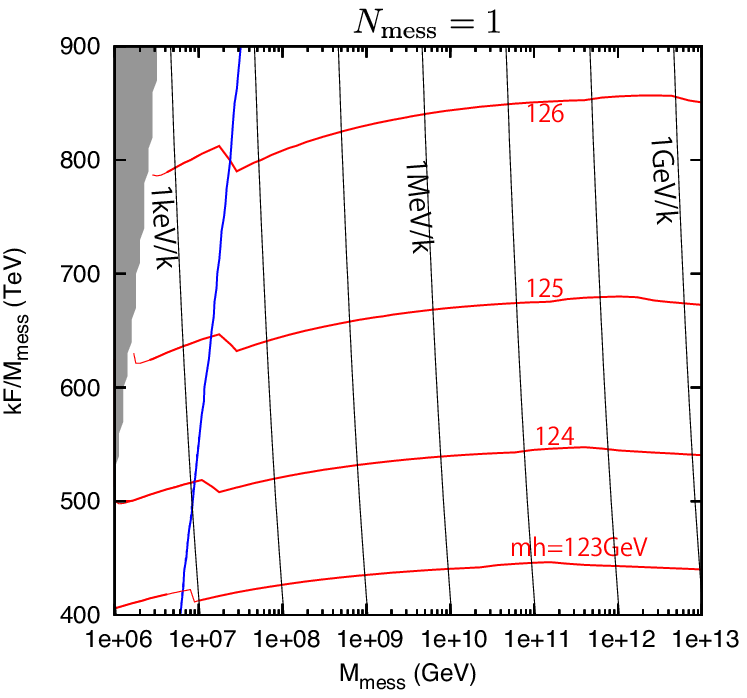}
\includegraphics[scale=0.95]{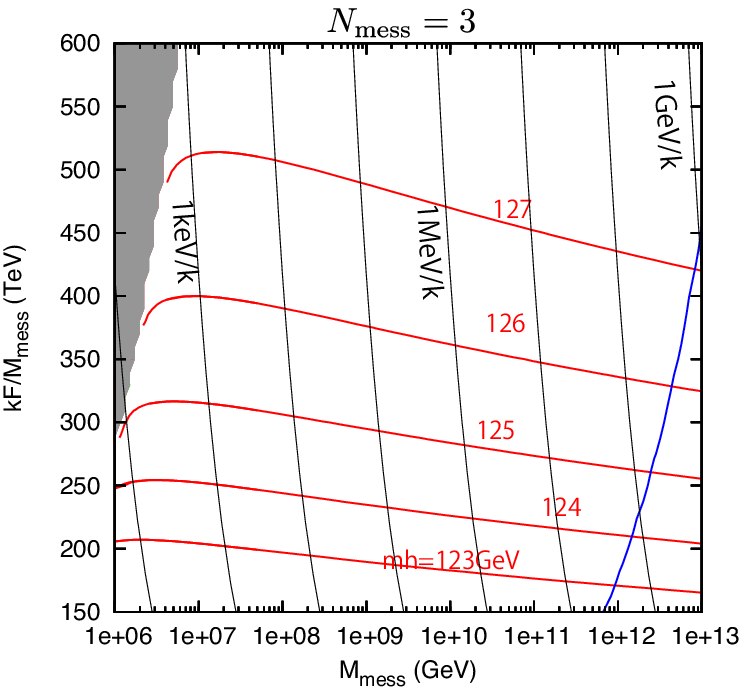}
\caption{\sl \small Contours of the gravitino mass decided by $k$. 
}
\label{fig:gravitino}
\end{center}
\end{figure}

\begin{figure}
\begin{center}
\includegraphics[scale=0.95]{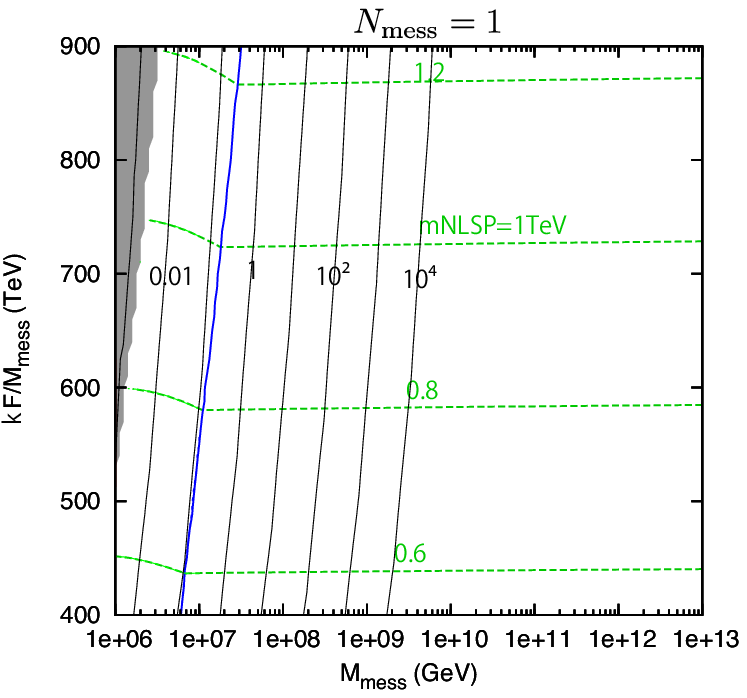}
\includegraphics[scale=0.95]{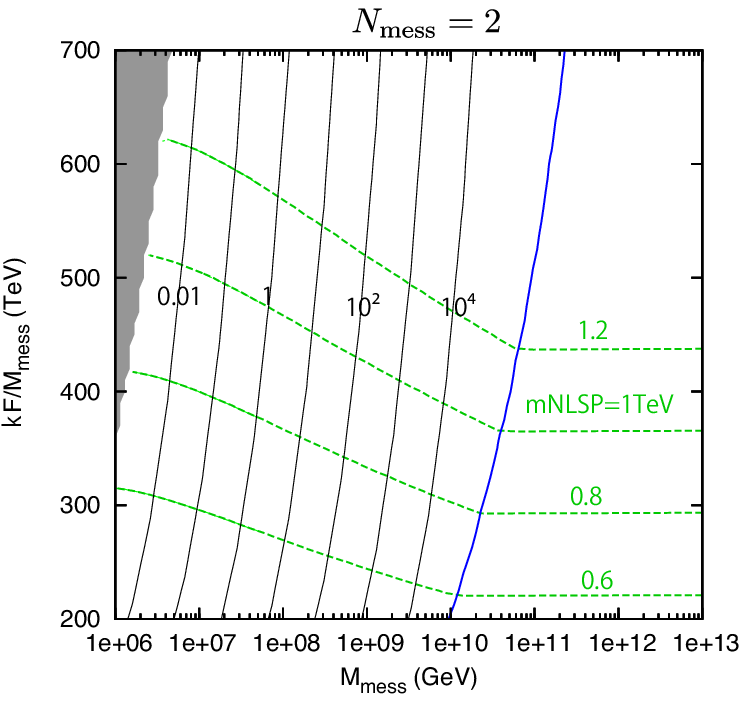}
\includegraphics[scale=0.95]{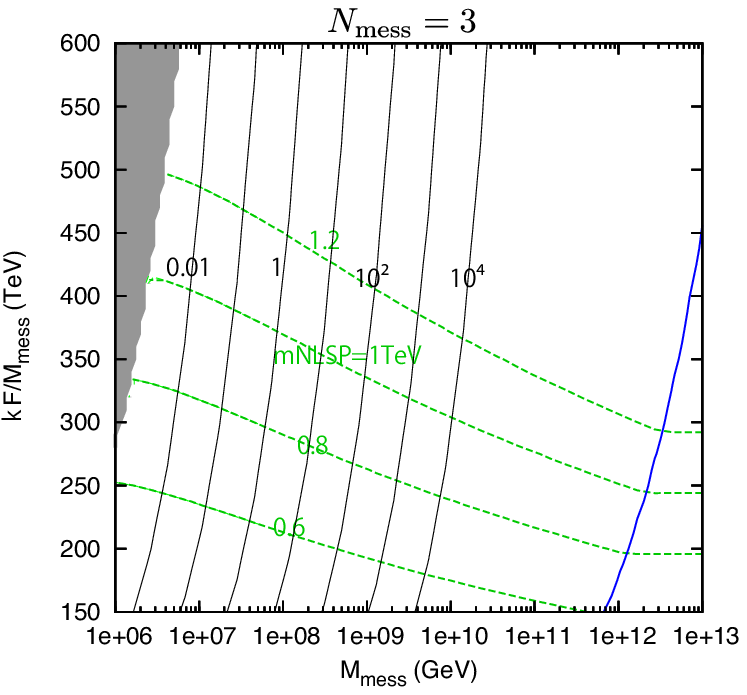}
\includegraphics[scale=0.95]{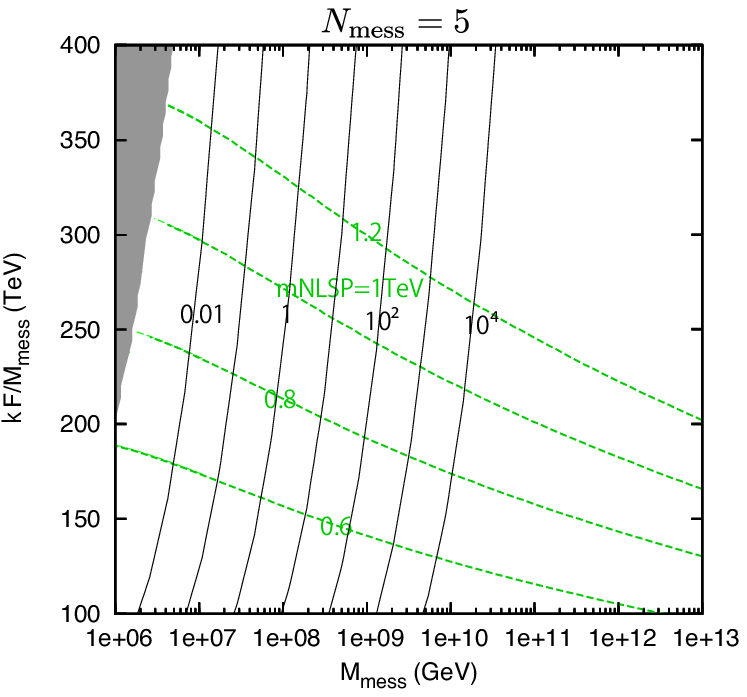}
\caption{\sl\small The decay length of the NLSP ($c\tau_{\rm NLSP} (0.1/k)^2$) in the unit of meter.
}
\label{fig:decay}
\end{center}
\end{figure}

To search for the  SUSY signals at the LHC, the decay length of the NLSP is very important, which depends on the 
gravitino mass (see the appendix\,\ref{sec:lifetime}).
In Fig.~\ref{fig:gravitino}, we show the gravitino mass (decided by $k$) for given parameters.
The figures show that the gravitino is bounded from below, $m_{3/2} \gtrsim 0.1-1$\,keV in most parameter region. 
In Fig.~\ref{fig:decay}, the decay length of the lightest stau ($c\tau_{\rm stau}$) are shown in the unit of $m$.
At the 14 TeV run of the LHC experiment, it will be able to search for the stable stau inside the detector 
with mass below about $m_{\rm \tau}\lesssim 1.0$\,TeV by combining 
the measurements of the ionizing energy loss rate ($dE/dx$) and the time-of-flight~\cite{cms_stau}. 
Therefore, the MDGM model can be tested at the LHC when the NLSP is the stable stau.

\begin{figure}[t]
\begin{center}
\includegraphics[scale=0.95]{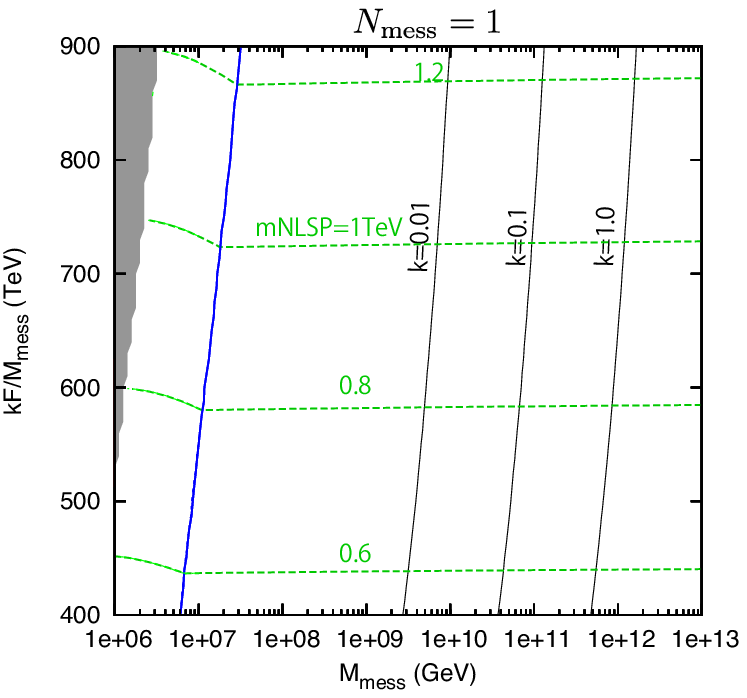}
\includegraphics[scale=0.95]{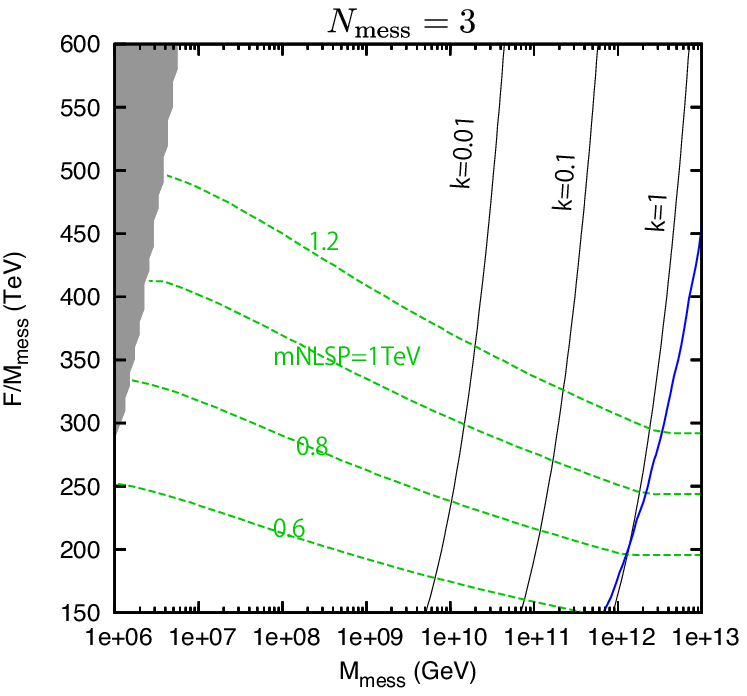}
\caption{\sl\small Upper-bound for the messenger scale from the electron EDM ($|d_e| < 8.7 \times 10^{-29} e {\rm cm}$). Here, we assume that the relevant phase arises from the Higgs $B$-term as  $B=B_{\rm GMSB} + i |m_{3/2}|$.
}
\label{fig:edm}
\end{center}
\end{figure}

Before closing this section, let us comment on the CP violation from the supergravity mediated effects.
As we have discussed in the previous section, all the phased in the MDGM model can be rotated away.
Therefore, no additional source of the CP violation is introduced in MDGM, except for the 
supergravity mediated $O(m_{3/2})$  corrections to soft mass parameters. 
Since a larger gravitino mass is required for a larger messenger scale,
the CP violation from the $O(m_{3/2})$ corrections may become important for a large messenger scale. 
For instance, the electric dipole moment (EDM) of the electron can exceed the experimental bound\,\cite{Moroi:2011fi}.
In Fig.~\ref{fig:edm}, the upper-bounds on the messenger scales 
from the EDM constraint, $|d_e| < 8.7 \times 10^{-29}e$cm~\cite{electron_edm}, 
are shown for $k=0.01, 0.1$ and $1.0$. 
Here we have assumed that the CP phase arising from the supergravity effect is given by ${\arg}(B_{\rm GMSB} - i|m_{3/2}|)$.
For instance, for $k=0.1$, the region with a large messenger scale,
$M_{\rm mess} \gtrsim 10^{11}$ GeV may be excluded by the electron EDM constraint.
The corresponding gravitino mass is $m_{3/2} \sim $ 0.1\,GeV.

\section{Discussions and Conclusions}
In this paper, we have reconsidered the models with gauge mediation in view of the minimality.
As a result, we have arrived at a very simple model, the MDGM, which is free from the SUSY FCNC problems
and the SUSY CP problems.
Interestingly, we have found parameter space  with the Higgs boson mass at around $126$\,GeV 
which can be tested at the 14\,TeV run of the LHC through the stable stau search.%
\footnote{
This prediction can be contrasted with the predictions in pure gravity mediation 
model\,\cite{Ibe:2011aa,Ibe:2012hu}/minimal-split SUSY model\,\cite{ArkaniHamed:2012gw}
which are also favored in view of minimality, where the model can be tested 
through the gaugino searches\,\cite{Bhattacherjee:2012ed}.
}

As we have discussed, the MDGM model predicts the gravitino mass $m_{3/2}\gtrsim 1$\,keV
in  most of the parameter space. 
In this mass region, the gravitino is most likely stable or long lived, and hence, it
is a good candidate of dark matter (DM). 
However, its thermal abundance exceeds the observed density of the DM,
which requires some mechanism to dilute the dark matter density. 
Notably, in the MDGM model, we already have a candidate of the source of the entropy,
the messenger fields.
As proposed in Ref.\,\cite{Fujii:2002fv}, the energy density of the messenger fields can dominate
over the radiation when it is long lived, and produce a lot of entropy at their decay.
With the help of the entropy production  mechanism we  found a large parameter space where the observed 
DM density is explained by the gravitino of mass in the range of about $1\,$\,keV--$1$\,GeV (see the appendix\,\ref{sec:entropy}).

If we embed the present model in string theories, there arises an
intriguing candidate for DM besides the gravitino, that is the string
moduli whose masses are of order of the gravitino mass. For
$m_{3/2}=1-100$ keV the moduli field is long lived and a candidate for the DM
\cite{Kawasaki}. 
The moduli decay into two photons with the Planck suppressed operator 
which can be accessible in  cosmic X-ray telescopes
for $m_{3/2}=1-100$ keV \cite{Kusenko, Kawasaki}.
It is very exciting that two groups analyzing  the X-ray data of the many galaxy
clusters have recently reported unidentified line signals  at $3.5$\,keV\,\cite{Bulbul:2014sua,Boyarsky:2014jta}.%
\footnote{
See for example, Ref.\,\cite{Ishida:2014dlp,Finkbeiner:2014sja,Higaki:2014zua,Jaeckel:2014qea,Lee:2014xua, xray_0306}
for some recent ideas to explain the line signals by DM.
}
We will discuss more details in a future publication\,\cite{HIYY}.

\section*{Acknowledgment}

This work was supported by JSPS KAKENHI Grant 
No.~22244021 (K.H., T.T.Y), and also by World Premier International Research Center Initiative
(WPI Initiative), MEXT, Japan. The work of NY is supported in part by JSPS Research Fellowships for Young Scientists.

\appendix

\section{The Lifetime of the NLSP}
\label{sec:lifetime}
In the models with gauge mediation, the decay length of the NLSP is important 
for the LHC phenomenology.
In this appendix, we summarize the decay rate of the NLSP into a gravitino.

The decay rate of the NLSP stau is given by
\begin{align}
(c\tau_{\tilde{\tau}})^{-1}
&\simeq\Gamma(\tilde{\tau}\to \tilde{G}+\tau)
\\
&=
\frac{1}{48\pi M_P^2}
\frac{m_{\tilde{\tau}}^5}{m_{3/2}^2}
\left(
1-\frac{m_{3/2}^2}{m_{\tilde{\tau}}^2}
\right)^4
\\
&\simeq (1.8~\text{m})^{-1}
\left(\frac{m_{\tilde{\tau}}}{1~\text{TeV}}\right)^5
\left(\frac{100~\text{keV}}{m_{3/2}}\right)^2\ ,
\end{align}
where we have assumed $m_{\tilde{\tau}}\gg m_{\tau}$.

The decay rate of the NLSP neutralino is given by
\begin{align}
(c\tau_{\tilde{\chi}})^{-1}
&\simeq 
\Gamma(\tilde{\chi}\to \tilde{G}+\gamma)
+
\Gamma(\tilde{\chi}\to \tilde{G}+Z) \ ,
\end{align}
where~\cite{Feng:2004mt}
\begin{align}
\Gamma(\tilde{\chi}\to \tilde{G}+\gamma)
&=
\frac{1}{48\pi M_P^2}
\frac{m_{\tilde{\chi}}^5}{m_{3/2}^2}
\left| N_{1\tilde{B}}\cos\theta_W + N_{1\tilde{W}}\sin\theta_W \right|^2
f\left(\frac{m_{3/2}}{m_{\tilde{\chi}}}\right)\ ,
\\
\Gamma(\tilde{\chi}\to \tilde{G}+Z)
&=
\frac{1}{48\pi M_P^2}
\frac{m_{\tilde{\chi}}^5}{m_{3/2}^2}
\left| -N_{1\tilde{B}}\sin\theta_W + N_{1\tilde{W}}\cos\theta_W \right|^2
g\left(\frac{m_{3/2}}{m_{\tilde{\chi}}},
\frac{m_Z}{m_{\tilde{\chi}}}\right)\ ,
\end{align}
with $N_{1\tilde{B}}$ and $N_{1\tilde{W}}$ being neutralino
mixing angles and 
\begin{align}
f\left(r_{3/2}\right) &= 
\left(1-r_{3/2}^2\right)^3
\left(1+3r_{3/2}^2\right)\ ,
\\
g\left(r_{3/2},r_Z\right)
&=
\sqrt{1-2\left(r_{3/2}^2 + r_Z^2\right) + \left(r_{3/2}^2-r_Z^2\right)^2}
\nonumber\\
&\times
\left[
\left(1-r_{3/2}^2\right)^3
\left(1+3r_{3/2}^2\right)
-
r_Z^2
\left(
3+r_{3/2}^3(-12+r_{3/2}) + r_Z^4 - r_Z^2(3-r_{3/2}^2)
\right)
\right]\ ,
\end{align}
For $m_{\tilde{\chi}}\gg m_Z, m_{3/2}$ and $N_{\tilde{B}1}\simeq 1$,
it is given by
\begin{align}
(c\tau_{\tilde{\chi}})^{-1}
&\simeq (1.8~\text{m})^{-1}
\left(\frac{m_{\tilde{\chi}}}{1~\text{TeV}}\right)^5
\left(\frac{100~\text{keV}}{m_{3/2}}\right)^2\ ,
\end{align}

\section{Gravitino Dark Matter in MDGM}\label{sec:entropy}

Here, we show that the LSP gravitino in the MDGM model
is a viable and natural dark matter candidate.
Let us consider, as an example, the case of $N_5=3$.
As shown in section\,\ref{sec:predictions}
the Higgs mass requires that $kF/M_\text{mess}\simeq 300-400$ TeV.
There, the NLSP is the stau with a mass of $m_{\tilde{\tau}}\simeq 0.8-1.2~\text{TeV}$,
and the gluino mass is about $m_{\tilde{g}}\simeq 5~\text{TeV}$.
 The gravitino mass is then given by
\begin{align}
m_{3/2} &\simeq 0.83~\text{MeV}
\left(\frac{M_{\text{mess}}}{10^9~\text{GeV}}\right)
\times
\left(\frac{0.1}{k}\right)
\left(\frac{kF/M_\text{mess}}{350~\text{TeV}}\right)\,.
\label{eq:m32}
\end{align}
If there is no late--time entropy production, the gravitino abundance in the present universe is determined by the reheating temperature $T_R$~\cite{Omega32},
$\Omega_{3/2}h^2\simeq 0.3\times
(T_R/10^7~\text{GeV})\allowbreak
(10~\text{GeV}/m_{3/2})
(m_{\tilde{g}}/5~\text{TeV})^2$.
For instance, for $(M_\text{mess}, (kF/M_\text{mess}), k)\simeq 
(10^{12}~\text{GeV},\allowbreak 300~\text{TeV},\allowbreak 0.01)$, 
the gravitino mass becomes $m_{3/2}\simeq 7~\text{GeV}$ and it can explain the dark matter density for $T_R\simeq {\cal O}(10^6)~\text{GeV}$.
However, such a large gravitino mass may cause a too large CP violation (see Fig.\,\ref{fig:edm}).
In addition, $T_R\simeq 10^6~\text{GeV}$ is too low for a successful thermal leptogenesis~\cite{Fukugita:1986hr}, which is one of the most attractive baryogenesis mechanisms.

Interestingly, another viable scenario opens up if the reheating temperature becomes 
higher than the messenger mass, $T_R> M_\text{mess}$~\cite{Fujii:2002fv}.
In the present scenario, both the gravitino and the messenger fields
become in thermal equilibrium for $T_R> M_\text{mess}$.\footnote{This is the case as far as $M_{\text{mess}}\ll 10^{15}~\text{GeV}\times (k/0.1)^2$. If this inequality is not satisfied, gravitinos are not necessarily in thermal equilibrium for $T_R>M_{\text{mess}}$.
Gravitino dark matter scenarios in such a case are discussed in Refs.~\cite{ref_small_k}.}
The resultant gravitino abundance, $\Omega_{3/2}^{\text{eq}}h^2\simeq 5\times 10^3~(m_{3/2}/10~\text{MeV})$, would by far  exceed the observed dark matter density,
if there is no late-time entropy production.
However, in the minimal gauge mediation, there is a natural mechanism to dilute this gravitino abundance by right amount, by the decay of a metastable messenger field.
We assume that the following mixing term between the MSSM and the messenger fields
is induced by the R-symmetry breaking constant term $W_0$ in the superpotential~\cite{Fujii:2002fv}:
\begin{align}
\delta W = f_i \frac{{W_0}}{M_P^2} \Psi {\bf \bar{5}}_i
= f_i m_{3/2} \Psi {\bf \bar{5}}_i\,
\end{align}
where ${\bf \bar{5}}_i$ is the MSSM multiplet, $f_i$ are constants of order unity,
and $i$ denotes the generation index.
Then, the lightest messenger field, which is the scalar component of a weak doublet,
becomes long-lived and decays into Higgsino and SM lepton through this small mixing  with a rate;
\begin{align}
\Gamma_{\text{mess}} &\simeq
\frac{1}{8\pi}
\left(\frac{m_\tau}{v\cos\beta}\right)^2
\left(\frac{f_3 m_{3/2}}{M_{\text{mess}}}\right)^2 
M_{\text{mess}}
\\
&\simeq (6\times 10^{-9}\text{sec})^{-1}
f_3^2
\left(\frac{\tan\beta}{50}\right)^2
\left(\frac{m_{3/2}}{10~\text{MeV}}\right)^2
\left(\frac{10^{10}\text{GeV}}{M_\text{mess}}\right)
\,,
\end{align}
where $v\simeq 174$ GeV is the Higgs VEV,
and we have assumed that the decay into the third generation is dominant.
Thus, the messenger decays before the 
big bang nucleosynthesis as far as 
$f_3\gtrsim {\cal O}(10^{-4})(10~\text{MeV}/m_{3/2})(M_\text{mess}/10^{10}\text{GeV})^{1/2}$.  
The thermal relic abundance of the messenger field is given by $Y_{\text{mess}}=n_{\text{mess}}/s = 3.7\times 10^{-10}
(M_{\text{mess}}/10^6\text{GeV})$~\cite{Ymess,Fujii:2002fv}.
The energy density of the messenger field dominates the universe before its decay for $M_{\text{mess}}Y_{\text{mess}}\gtrsim T_d$, where $T_d\simeq (g_*/10)^{-1/4} \sqrt{M_P \Gamma_{\text{mess}}}$ with $g_*$ being the effective degrees of freedom at $T=T_d$.
 This condition is equivalent to
\begin{align}
\Delta &\equiv \frac{4}{3}
\frac{M_{\text{mess}}Y_{\text{mess}}}{T_d}
\nonumber\\
&\simeq 
3\times 10^3
\frac{1}{f_3}
\left(\frac{g_*}{10}\right)^{1/4}
\left(\frac{50}{\tan\beta}\right)
\left(\frac{M_{\text{mess}}}{10^{10}\text{GeV}}\right)^{5/2}
\left(\frac{10~\text{MeV}}{m_{3/2}}\right)
\; > \; 1\,.
\label{eq:dom}
\end{align}
If this is satisfied, the final gravitino abundance is given by
\begin{align}
\Omega_{3/2}h^2 &\simeq 
\frac{1}{\Delta}\Omega_{3/2}^{\text{eq}}h^2
\nonumber \\
&\simeq 0.16
\left(\frac{f_3}{0.01}\right)
\left(\frac{10}{g_*}\right)^{1/4}
\left(\frac{\tan\beta}{50}\right)
\left(\frac{m_{3/2}}{10~\text{MeV}}\right)^2
\left(\frac{10^{10}~\text{GeV}}{M_\text{mess}}\right)^{5/2}
\end{align}
Therefore, the LSP gravitino in the minimal gauge mediation  can explain the present dark matter density 
in a wide range the parameter space consistent with the Higgs mass $m_h\simeq 126$~GeV (see Fig.\,\ref{fig:gravitino}),
with moderate values of $k$ and $f_3$.
We emphasize that the gravitino abundance is independent of the reheating temperature
as far as $T_R>M_{\text{mess}}$, and that the thermal leptogenesis works successfully~\cite{Fujii:2002fv}.

\end{document}